\pdfoutput=1

\documentclass[usenatbib,useAMS,twocolumn,fleqn]{mnras}
\usepackage[dvipdfmx]{graphicx}
\usepackage[dvipdfmx]{color}
\usepackage{amsmath,amssymb}
\usepackage{epstopdf}
\usepackage{grffile}
\usepackage{times}
\usepackage{url}
\usepackage{bm}
\usepackage{xcolor}

\usepackage[dvipdfmx]{graphicx}	
\usepackage{amsmath}	
\usepackage{amssymb}	
\usepackage{color}
\usepackage{dcolumn}
\usepackage{bm}
\usepackage{comment}

\newcommand{\beq}{\begin{equation}}
\newcommand{\beqa}{\begin{eqnarray}}
\newcommand{\eeq}{\end{equation}}
\newcommand{\eeqa}{\end{eqnarray}}

\newcommand{\simgt}{\lower.5ex\hbox{$\; \buildrel > \over \sim \;$}}
\newcommand{\simlt}{\lower.5ex\hbox{$\; \buildrel < \over \sim \;$}}

\newcommand{\bd}[1]{\mbox{\boldmath $#1$}}

\RequirePackage{lineno}

\def\apj{{ApJ}}                 
\def\apjl{{ApJ}}                
\def\apjs{{ApJS}}               
\def\ao{{Appl.~Opt.}}           
\def\aap{{A\&A}}                

\def\mnras{MNRAS}             
\def\prd{{Phys.~Rev.~D}}        
\def\pasj{{PASJ}}               
\def\nat{{Nature}}              

\title[Weak lensing peak counts]{
Statistical connection of peak counts to 
power spectrum and moments in weak lensing field
}

\author[M.~Shirasaki]
{Masato Shirasaki
\thanks{E-mail:masato.shirasaki@nao.ac.jp},
\\
Division of Theoretical Astronomy, National Astronomical Observatory of Japan, 2-21-1 Osawa, Mitaka, Tokyo 181-8588, Japan
}

\date{Accepted XXX. Received YYY; in original form 3/3/2016}

\pubyear{2016}

\begin{document}
\label{firstpage}
\pagerange{\pageref{firstpage}--\pageref{lastpage}}
\maketitle

\begin{abstract}
The number density of local maxima of 
weak lensing field, referred to as weak-lensing peak counts, 
can be used as a cosmological probe.
However, its relevant cosmological information is still unclear.
We study the relationship between the peak counts 
and other statistics in weak lensing field 
by using 1000 ray-tracing simulations.
We construct a local transformation of lensing field $\cal K$
to a new Gaussian field $y$, named local-Gaussianized transformation.
We calibrate the transformation with numerical simulations 
so that the one-point distribution and the power spectrum 
of $\cal K$ can be reproduced from a single Gaussian field $y$ 
and monotonic relation between $y$ and $\cal K$.
Therefore, the correct information of two-point clustering and 
any order of moments in weak lensing field
should be preserved under local-Gaussianized transformation.
We then examine if local-Gaussianized transformation 
can predict weak-lensing peak counts in simulations.
The local-Gaussianized transformation is insufficient to
explain weak-lensing peak counts in the absence of shape noise.
The prediction by local-Gaussianized transformation
underestimates the simulated peak counts with a level of
$\sim20-30\%$ over a wide range of peak heights.
Local-Gaussianized transformation can predict the weak-lensing peak counts with 
a $\sim10\%$ accuracy in the presence of shape noise.
Our analyses suggest that the cosmological information beyond power spectrum 
and its moments would be necessary to predict the weak-lensing peak counts 
with a percent-level accuracy,
which is an expected statistical uncertainty in upcoming wide-field galaxy surveys.
\end{abstract}

\section{INTRODUCTION}

Gravitational lensing is caused by 
bending of the light emitted from a distant source, 
which is a prediction by General Relativity. 
Since the exact amount of bending depends on projected mass along a line of sight, 
gravitational lensing is now recognised as 
direct and very promising probe of the matter distribution in the Universe. 
In general, foreground matter causes small distortions 
of the images of distant background galaxies, 
and these small distortions contain rich information 
on the foreground matter distribution and its growth over time, 
which are sensitive to the nature of dark energy and dark matter
\citep[e.g.,][for a review]{2010GReGr..42.2177H, 2015RPPh...78h6901K}.
In order to extract cosmological information from image distortion induced by gravitational lensing,
various statistical methods have been proposed in the literature.
Mapping matter distribution on a continuous sky is a key, 
basic method among them
\citep[e.g.,][]{2007Natur.445..286M, 2013MNRAS.433.3373V, 2015PhRvL.115e1301C}.
Non-Gaussian features in the reconstructed
density field are produced by non-linear gravitational growth and 
can not be extracted by means of the conventional statistics of 
cosmic shear such as two-point correlation function or power spectrum
\citep[e.g.,][]{2003ApJ...592..664P}.

The statistics of local maxima found in a reconstructed lensing map
have cosmological information originated from massive dark matter haloes.
The height of peaks are commonly normalised by 
the noise variance of a reconstructed map $\sigma$.
The local maxima with the height of $\simgt3\sigma$ significance are
expected to be caused by high density regions in the Universe 
that correspond to massive collapsed objects such as clusters of galaxies
\citep[e.g.,][]{2004MNRAS.350..893H, 2005ApJ...624...59H, 2005A&A...442..851M, 
2012MNRAS.423.1711M, 2012MNRAS.425.2287H}.
The one-to-one correspondence between clusters and lensing peaks 
has been confirmed by recent galaxy imaging surveys \citep[e.g.,][]{2007ApJ...669..714M, 2007A&A...462..875S, 2012ApJ...748...56S, 2015ApJ...807...22M}.
Therefore, the number count of high-significance peaks
can be associated with the abundance of galaxy clusters that is highly
sensitive to growth of matter density perturbations \citep[e.g.,][]{1992ApJ...386L..33L}.
The peak counts can be used to place interesting constraints 
on the matter density perturbation in the Universe
\citep[e.g.,][]{2006PhRvD..73l3525M, 2010MNRAS.402.1049D}, 
the nature of dark energy
\citep[e.g.,][]{2011ApJ...728L..13M}, 
the statistical property of initial density perturbations
\citep[e.g.,][]{2012MNRAS.426.2870H}
and the modification of General Relativity 
\citep[e.g.,][]{2016PASJ...68....4S}.
Several observational constraints also have been obtained in the recent lensing surveys
\citep[e.g.,][]{2015MNRAS.450.2888L, 2015PASJ...67...34H, 2016PhRvL.117e1101L}.

Numerical simulations suggest 
that cosmological information in the number density of lensing peaks
can be carried by not only high-significance peaks but also
the peaks with $\sim1-2\sigma$ significance
\citep{2010PhRvD..81d3519K}.
We call such peaks with intermediate height as medium peaks.
There exist observational studies to constrain on cosmological models
by counting medium peaks 
\citep[e.g.,][]{2015PhRvD..91f3507L, 2016MNRAS.463.3653K}.
Since the medium peaks have similar heights 
to the noise variance by definition, 
it is not straightforward to infer
their relevant cosmological information.
Fisher analyses with a large set of numerical simulations
clearly show that counting medium peaks can improve the cosmological
constraints by weak-lensing power spectrum alone \citep{2013ApJ...774...49B}.
The improvement of cosmological constraints by weak-lensing peak counts has 
been demonstrated with actual data \citep{2015PhRvD..91f3507L}.
Thus, medium peaks are expected to contain non-Gaussian information
but it is uncertain which levels of non-Gaussianity are important.
Detailed comparisons between the angular position of medium peaks 
and dark matter halos along a line of sight 
show that medium peaks can be generated by
superposition of large-scale structures and multiple dark matter halos, 
whereas the intrinsic ellipticities of sources can further compromise the correlation between
medium peaks and dark matter halos \citep{2011PhRvD..84d3529Y}.
\citet{2016PhRvD..94d3533L} perform similar analysis with actual data set to show the angular correlation of medium peaks with galaxies on sky, 
suggesting that the origin of medium peaks can not be explained 
by observational noise alone.
The number count of medium peaks is also found to be 
insensitive to the internal structure of dark matter halos 
\citep{2013PhRvD..87b3511Y} 
and the details of baryonic physics 
\citep{2015ApJ...806..186O}.
Although halo-based approaches have been developed to 
predict the peak counts over a wide range of
peak heights \citep[e.g.][]{2015A&A...576A..24L, 2016arXiv160903973Z},
it is still insufficient to make a prediction with a few percent accuracy, 
corresponding to the statistical uncertainty in upcoming lensing surveys.

To infer the cosmological information in weak-lensing peak counts, 
we study the relation of cosmological information 
in weak-lensing peak counts and other lensing statistics.
For comparison, we consider the power spectrum and its moments of weak lensing field.
We assume a simple monotonic relation between a reconstructed lensing map
and a new Gaussian field, named local-Gaussianized transformation
\citep[also see][for the related work]{2016arXiv160705007Y}.
We construct such relationship by a local mapping and 
calibrate that the one-point probability distribution function (PDF) and
the power spectrum of simulated lensing field can be reproduced with 
a $\sim5\%$ accuracy.
Thus, after calibrations, our model based on local-Gaussianized transformation
should include appropriate cosmological information of two-point clustering
and moments in calculation of expected peak counts.
By comparing such theoretical models and simulated peak counts,
we examine if the peak counts can be fully explained by
cosmological information of power spectrum and one-point PDF of lensing field.

The paper is organised as follows.
In Section~\ref{sec:wl}, we briefly describe 
the basics of weak gravitational lensing.
There, we also introduce a general expression of weak-lensing peak counts and
a local transformation of weak lensing field to a Gaussian field.
We also summarise our ray-tracing simulations and how to construct
a local-Gaussianized transformation from these simulations in Section~\ref{sec:sim}.
By using our theoretical model and a large set of ray-tracing simulations,
we examine if peak counts can include the cosmological information beyond
the two-point clustering and moments in Section~\ref{sec:res}.
Conclusions and discussions are summarised in Section~\ref{sec:con}.

\section{WEAK GRAVITATIONAL LENSING}\label{sec:wl}

\subsection{\label{subsec:basics}Basics}
We first summarise the basics of gravitational lensing 
induced by large-scale structure.
Weak gravitational lensing effect is usually characterised by
the distortion of image of a source object by the 
following 2D matrix:
\beqa
A_{ij} = \frac{\partial \beta_{i}}{\partial \theta_{j}}
           \equiv \left(
\begin{array}{cc}
1-\kappa -\gamma_{1} & -\gamma_{2}  \\
-\gamma_{2} & 1-\kappa+\gamma_{1} \\
\end{array}
\right), \label{distortion_tensor}
\eeqa
where we denote the observed position of a source object 
as $\bd{\theta}$ and the true position as $\bd{\beta}$,
$\kappa$ is the convergence, and $\gamma$ is the shear.
In the weak lensing regime (i.e., $\kappa, \gamma \ll 1$), 
one can express the weak lensing convergence field as the integral of matter overdensity field $\delta({\bm x})$ along a line of sight
\citep{2001PhR...340..291B},
\beqa
\kappa(\bd{\theta},\chi_{\rm source}) &=& 
\frac{3}{2}\left(\frac{H_{0}}{c}\right)^2 \Omega_{\rm m0}
\int _{0}^{\chi}{\rm d}\chi \, \nonumber \\
&&\times
\frac{r(\chi_{\rm source}-\chi)r(\chi)}{r(\chi_{\rm source})} \,
\frac{\delta[r(\chi)\bd{\theta},\chi]}{a(\chi)}, \label{eq:kappa_delta}
\eeqa
where 
$\chi$ is the comoving distance,
$\chi_{\rm source}$ is the comoving distance to a source,
$r(\chi)$ is the angular diameter distance,
$H_{0}$ is the present-day Hubble constant
and $\Omega_{\rm m0}$ represents the matter density parameter
at present.
Throughout this paper, 
we assume that source galaxies locate at single plane with the redshift 
of $z_{\rm source} = 1$ for simplicity.

\subsubsection*{Power spectrum}
The power spectrum is one of the basic statistics in modern cosmology 
\citep[e.g.,][]{2012MNRAS.427.3435A, 2015arXiv150702704P, Becker:2015ilr}.
It is defined as the two-point correlation in Fourier space: 
\begin{eqnarray}
\langle \tilde{\kappa}(\bm{\ell}_1) \tilde{\kappa}(\bm{\ell}_2) \rangle 
= (2\pi)^2 \delta_D (\bm{\ell}_{1}+\bm{\ell}_{2})P_\kappa (\ell_1),
\end{eqnarray}
where $\delta_{D}(\bm{x})$ is the Dirac delta function and 
the multipole $\ell$ is related to the angular scale through 
$\theta=\pi/\ell$. 
By using the Limber approximation \citep{Limber:1954zz,Kaiser:1991qi}
and Eq.~(\ref{eq:kappa_delta}), 
one can express the convergence power spectrum as 
\beqa
P_{\kappa}(\ell) = \int_{0}^{\chi_{\rm source}} {\rm d}\chi \frac{W(\chi)^2}{r(\chi)^2} 
P_{\delta}\left(k=\frac{\ell}{r(\chi)},z(\chi)\right)
\label{eq:kappa_power},
\eeqa
where $P_{\delta}(k)$ represents the three dimensional matter power spectrum, 
and $W(\chi)$ is the lensing weight function defined as
\beqa
W(\chi) = \frac{3}{2}\left(\frac{H_{0}}{c}\right)^2 
\Omega_{\rm m0} \frac{r(\chi_{\rm source}-\chi)r(\chi)}{r(\chi_{\rm source})}(1+z(\chi)).
\eeqa
 
\subsubsection*{Reconstruction of smoothed convergence}
In actual observations, 
one usually start with the cosmic shear
instead of the convergence field. 
The reconstruction of smoothed convergence is 
commonly based on the smoothed map of cosmic shear.
Let us first define the smoothed convergence map as
\beqa
{\cal K}(\bd{\theta}) = 
\int {\rm d}^2 \bd{\phi}\, \kappa(\bd{\theta}-\bd{\phi}) 
U (\bd{\phi}),
\eeqa
where $U$ is the filter function to be specified below.
We can calculate the same quantity by smoothing 
the shear field $\gamma$ as
\beqa
{\cal K} (\bd{\theta}) = \int {\rm d}^2 \bd{\phi} \ \gamma_{+}(\bd{\phi}:\bd{\theta}) Q_{+}(\bd{\phi}), \label{eq:ksm}
\eeqa
where $\gamma_{+}$ is the tangential component of the shear 
at position $\bd{\phi}$ relative to the point $\bd{\theta}$.
The filter function for the shear field $Q_{+}$ is related to $U$ by
\beqa
Q_{+}(\theta) = \int_{0}^{\theta} {\rm d}\theta^{\prime} \ \theta^{\prime} U(\theta^{\prime}) - U(\theta).
\label{eq:U_Q_fil}
\eeqa
We consider a filter function $Q_{+}$ that has a finite extent.
In such cases, one can write
\beqa
U(\theta) = 2\int_{\theta}^{\theta_{o}} {\rm d}\theta^{\prime} \ \frac{Q_{+}(\theta^{\prime})}{\theta^{\prime}} - Q_{+}(\theta),
\eeqa
where $\theta_{o}$ is the outer boundary of the filter function.
Note that the filter function $U$ should be compensated 
because the smoothed field $\cal K$ does not depend on 
undetermined constant \citep{1996MNRAS.283..837S}.

In the following, we consider the truncated Gaussian filter (for $U$):
\beqa
U(\theta) &=& \frac{1}{\pi \theta_{G}^{2}} 
\exp \left( -\frac{\theta^2}{\theta_{G}^2} \right) \nonumber \\
&&
\,\,\,\,\,\,\,\,
\,\,\,\,\,\,\,\,
\,\,\,\,\,\,\,\,
-\frac{1}{\pi \theta_{o}^2}\left[ 1-\exp \left(-\frac{\theta_{o}^2}{\theta_{G}^2} \right) \right], \label{eq:filter_kappa}\\
Q_{+}(\theta) &=& \frac{1}{\pi \theta^{2}}\left[ 1-\left(1+\frac{\theta^2}{\theta_{G}^2}\right)\exp\left(-\frac{\theta^2}{\theta_{G}^2}\right)\right],
\label{eq:filter_gamma}
\eeqa
for $\theta \leq \theta_{o}$ and $U = Q_{+} = 0$ elsewhere.
Throughout this paper, we set $\theta_{o} = 150$ arcmin
and study three cases of $\theta_{G} = 1.2, 2.4$ and $4.8$ arcmin.
Note that the choice of $\theta_{G} = 1.2$ arcmin 
is found to be an optimal smoothing scale for 
the detection of massive galaxy clusters using weak lensing for 
$z_{\rm source}$ = 1.0 \citep{2004MNRAS.350..893H}.

\subsection{\label{subsec:peak}Peak counts}
In this paper, we consider the number density of local maxima 
in smoothed convergence field $\cal K$.
The number density of extrema is expressed as
\citep{1986ApJ...304...15B, 1987MNRAS.226..655B}
\beqa
n_{\rm ext}({\bm \theta}) 
= \sum_{\rm ext} \delta_{D}({\bm \theta}-{\bm \theta}_{\rm ext})
= \left|\det{\bm \zeta}\right|\, \delta_{D}({\bm \eta}),
\eeqa
where $\eta_{i} = \nabla_{i} {\cal K}$, ${\zeta}_{ij} = \nabla_{i}\nabla_{j}{\cal K}$ and we use the relation that 
$\eta_{i}({\bm \theta}) = 
\zeta_{ij}(\theta_j-\theta_{{\rm ext}, j})$ 
around the position of extrema ${\bm \theta}_{\rm ext}$.
For a given multiplicative PDF
of ${\cal K}, {\bm \eta}$ and ${\bm \zeta}$,
the expected number density of local maxima with fixed peak height of $\alpha$
can be obtained by
\beqa
\frac{{\rm d}n_{\rm peak}}{{\rm d}\cal K}\biggr|_{{\cal K}=\alpha}
&=&
\int {\rm d}{\bm D} \, {\rm Prob}({\bm D})
\delta_{D}({\cal K} - \alpha) \, \left|\det{\bm \zeta}\right|\, \nonumber \\
&&
\,\,\,\,\,\,\,\,
\,\,\,\,\,\,\,\,
\,\,\,\,\,\,\,\,
\times
\delta_{D}({\bm \eta}) \, {\cal H}({\bm \lambda}),
\label{eq:peak_dens_wonoise}
\eeqa
where
${\bm D} = \{ {\cal K}, {\bm \eta}, {\bm \zeta} \}$,
$\lambda_{i}$ represents the eigen value of $-{\bm \zeta}$
and ${\cal H}(x)$ is the heaviside step function.

Eq.~(\ref{eq:peak_dens_wonoise}) is a general expression of 
the number density of local maxima for a given two-dimensional field.
It is more useful to decompose the observed ${\cal K}$ field into two contributions;
\beqa
{\cal K} = {\cal K}_{\rm cosmo} + {\cal N}, \label{eq:kappa_decomp}
\eeqa
where ${\cal K}_{\rm cosmo}$ represents the cosmological convergence from large-scale structure in the Universe 
and ${\cal N}$ is the noise convergence coming 
from intrinsic ellipticity of source galaxies known as shape noise.
On the assumption that the intrinsic ellipticities of two distant galaxies are uncorrelated each other, the statistical property of ${\cal N}$ should 
be characterised by a Gaussian distribution.
Then, the PDF of ${\bm D}_{N} = \{ {\cal N}, {\bm \eta}_{N}, {\bm \zeta}_{N} \}$
is given by
\beqa
{\rm Prob}({\bm D}_{N}) =
\frac{1}{(2\pi)^3 \sqrt{\det{\bm M}_{N}}} 
\exp\left(-\frac{1}{2}{\bm D}_{N}^{T} {\bm M}_{N}^{-1}{\bm D}_{N} \right), \label{eq:prob_noise}
\eeqa
where ${\bm M}_{N}$ is the correlation matrix of noise field.
The correlation matrix ${\bm M}_{N}$ takes the following forms: 
\beqa
\langle {\cal N}{\cal N} \rangle &=& \sigma_{{\cal N}0}^2, \\
\langle {\cal N}{\eta}_{N, i} \rangle &=& 0, \\
\langle {\cal N}{\zeta}_{N, ij} \rangle &=& -\frac{1}{2}\sigma_{{\cal N}1}^2 \delta_{ij},\\
\langle {\eta}_{N, i}{\eta}_{N, j} \rangle &=& \frac{1}{2}\sigma_{{\cal N}1}^2 \delta_{ij},\\
\langle {\eta}_{N, i}{\zeta}_{N, jk} \rangle &=& 0, \\
\langle {\zeta}_{N, ij}{\zeta}_{N, kl} \rangle &=& \frac{1}{8}\sigma_{{\cal N}2}^2 \left( \delta_{ij}\delta_{kl}+ \delta_{ik}\delta_{jl}+\delta_{il}\delta_{jk}\right),
\eeqa
where $\sigma_{{\cal N}i}^2$ represents 
the $i$-th moment of ${\cal N}$. 
It is calculated from the power spectrum of ${\cal N}$ as follows
\citep{2000MNRAS.313..524V};
\beqa
\sigma_{{\cal N}i}^2 = \frac{\sigma_{e}^2}{2n_{\rm gal}} 
\int_0^{\infty} \frac{{\rm d}^{2}\ell}{(2\pi)^{2}}\, \ell^{2i} \, \left|\tilde{U}(\ell)\right|^2,
\label{eq:noise_moment}
\eeqa
where $\sigma_{e}$ is the rms value of the intrinsic ellipticity of 
the source galaxies,
$n_{\rm gal}$ is the number density of galaxies,
and $\tilde{U}(\ell)$ represents the Fourier transform of Eq.~(\ref{eq:filter_kappa}).
When $\theta_{o}$ is set to be 150 arcmin in Eq.~(\ref{eq:filter_kappa}),
the functional form of $|\tilde{U}(\ell)|^2$ can be approximated as
\beqa
|\tilde{U}(\ell)|^2 
&=& \exp\left[-\frac{1}{2}\left(\frac{\ell}{\ell_G}\right)^2\right] +1.5
\nonumber \\
&&\times10^{-5} \left(\frac{\ell}{1000}\right)^{-3}
\left[1+\left(\frac{\ell}{1000}\right)^{2.5}\right],
\eeqa
where ${\ell_G}=1/\theta_{G}=3438 \, (\theta_G/1\, {\rm arcmin})^{-1}$.

By using Eqs.~(\ref{eq:kappa_decomp}) and (\ref{eq:prob_noise}),
one can find the following expression of ${\rm d}n_{\rm peak}/{\rm d}{\cal K}$
in the presence of shape noise instead of Eq.~(\ref{eq:peak_dens_wonoise});
\beqa
\frac{{\rm d}n_{\rm peak}}{{\rm d}\cal K}\biggr|_{{\cal K}=\alpha}
&=&
\int {\rm d}{\bm D} \, {\rm Prob}({\bm D}|{\bm D}_{\rm cosmo})
\int {\rm d}{\bm D}_{\rm cosmo}\, {\rm Prob}({\bm D}_{\rm cosmo}) \nonumber \\
&&\times
\delta_{D}({\cal K} - \alpha) \, \left|\det{\bm \zeta}\right|\,
\delta_{D}({\bm \eta}) \, {\cal H}({\bm \lambda}) \nonumber \\
&=&
\int {\rm d}{\bm D}_{N} \, {\rm Prob}({\bm D}_{N})
\int {\rm d}{\bm D}_{\rm cosmo}\, {\rm Prob}({\bm D}_{\rm cosmo}) \nonumber \\
&&\times
\delta_{D}({\cal K} - \alpha) \, \left|\det{\bm \zeta}\right|\,
\delta_{D}({\bm \eta}) \, {\cal H}({\bm \lambda})
\label{eq:peak_dens_wnoise},
\eeqa
where 
${\bm D}_{\rm cosmo} = \{ {\cal K}_{\rm cosmo}, {\bm \eta}_{\rm cosmo}, {\bm \zeta}_{\rm cosmo} \}$
and we note that 
${\rm d}{\bm D}\, {\rm Prob}({\bm D}|{\bm D}_{\rm cosmo})
={\rm d}{\bm D}_{N}\, {\rm Prob}({\bm D}_{N})$ where
${\bm D}_{N}={\bm D}-{\bm D}_{\rm cosmo}$ holds.
When ${\cal K}_{\rm cosmo}$ is assumed to be Gaussian random field,
the number density of local maxima in ${\cal K}$ field is expressed as
the following analytic formula \citep{1987MNRAS.226..655B}:
\beqa
\frac{{\rm d}n_{\rm peak}}{{\rm d}\cal K}\biggr|_{{\cal K}=\alpha}
&=& \frac{1}{2\pi\theta_{*}^2}\frac{1}{\sqrt{2\pi}\sigma_{0}}
\exp\left[-\frac{1}{2}\left(\frac{\alpha}{\sigma_{0}}\right)^2\right]
\nonumber \\
&&\times
G\left(\Gamma, \Gamma\frac{\alpha}{\sigma_{0}}\right),
\label{eq:gauss_peak} 
\\
G(\Gamma, x_{*}) 
&=& 
(x_{*}^2-\Gamma^2)
\Biggl[ 
1 - \frac{1}{2}
{\rm erfc}\left(\frac{x_{*}}{\sqrt{2(1-\Gamma^2)}}\right)
\Biggr] \nonumber \\
&&
+\frac{x_{*}(1-\Gamma^2)}{\sqrt{2\pi(1-\Gamma^2)}}
\exp\left(-\frac{x_{*}^2}{2(1-\Gamma^2)}\right) \nonumber \\
&&
+\frac{1}{\sqrt{3-2\Gamma^2}}\exp\left(-\frac{x_{*}^2}{3-2\Gamma^2}\right)
\Biggl[ 1 - \frac{1}{2} \nonumber \\
&&\times
{\rm erfc}\left(\frac{x_{*}}{\sqrt{2(1-\Gamma^2)(3-2\Gamma^2)}}\right)
\Biggr], 
\label{eq:peak_function_G} \\
\theta_{*} &=& \frac{\sqrt{2}\sigma_{1}}{\sigma_{2}},\, \, 
\Gamma = \frac{\sigma_{1}^2}{\sigma_{0}\sigma_{2}},
\eeqa
where
\beqa
\sigma_{i}^2 = 
\int_0^{\infty} \frac{{\rm d}^{2}\ell}{(2\pi)^{2}}\, \ell^{2i} \, \left|\tilde{U}(\ell)\right|^2
\left(\frac{\sigma_{e}^2}{2n_{\rm gal}} + P_{\kappa}(\ell) \right).
\label{eq:observed_moment}
\eeqa

\subsection{Local-Gaussianized transformation}

\subsubsection*{Assumption}
In Section~\ref{subsec:peak}, 
we derive a general expression of the number density of local maxima 
in smoothed convergence field and 
find that the multiplicative PDF of ${\cal K}$, 
its first- and second-derivatives in the absence of shape noise is 
necessary to predict the peak counts.
Here we develop a theoretical model of multiplicative PDF 
for cosmological convergence field ${\cal K}_{\rm cosmo}$
by assuming a transformation
\beqa
{\cal K}_{\rm cosmo}({\bm \theta}) = {\cal F}\left[y\left({\bm \theta}\right)\right], \label{eq:local_transform}
\eeqa
where $y$ represents a Gaussian field with unit variance\footnote{
The mean value of $y$ can be determined by zero mean of ${\cal K}_{\rm cosmo}$.}.
The functional form of $\cal F$ is determined by the one-point PDF 
of ${\cal K}_{\rm cosmo}$ through the following equation:
\beqa
\int_{{\cal K}_{\rm cosmo}}^{\infty} {\rm d}{\cal K}_{\rm cosmo}^{\prime}\, {\rm Prob}({\cal K}_{\rm cosmo}^{\prime}) = \frac{1}{2}{\rm erfc}\left(\frac{y}{\sqrt{2}}\right). \label{eq:set_local_relation}
\eeqa
The statistical property of Gaussian field $y$ is fully characterised
by its power spectrum $P_{y}(\ell)$.
When adopting Eq.~(\ref{eq:local_transform}), one expect that
the two-point correlation function of $y$ is related to that of 
smoothed convergence field $\cal K$ as
\beqa
\xi_{y}(\phi) &\equiv& 
\langle y({\bm \theta}) y({\bm \theta}+{\bm \phi})\rangle, \\
\xi_{\cal K}(\phi) &\equiv& 
\langle {\cal K}_{\rm cosmo}({\bm \theta}) {\cal K}_{\rm cosmo}({\bm \theta}+{\bm \phi})\rangle \nonumber \\
&=& 
\frac{1}{2\pi\sqrt{1-\xi_{y}(\phi)^2}}
\int_{-\infty}^{\infty} {\rm d}u_{1} \int_{-\infty}^{\infty} {\rm d}u_{2}\, \nonumber \\
&&\times
\exp\left(-\frac{u_{1}^2}{2(1-\xi_{y}(\phi)^2)}\right)
\exp\left(-\frac{u_{2}^2}{2} \right) \nonumber \\
&&\times
{\cal F}(u_{1}+\xi_{y}(\phi)u_{2}){\cal F}(u_2). \label{eq:xi_relation}
\eeqa
Eq.~(\ref{eq:xi_relation}) defines the relation between $\xi_{\cal K}$ and $\xi_{y}$.
In the standard $\Lambda$CDM cosmology, the relation from 
Eq.~(\ref{eq:xi_relation}) is found to be expressed as a monotonic function of
$\xi_{y} = {\cal X}(\xi_{\cal K})$.
Therefore, one can obtain the power spectrum of $y$ by 
using Fourier transform and the relation from Eq.~(\ref{eq:xi_relation})
as follows
\beqa
P_{y}(\ell) &=& 
\int {\rm d}^{2}\theta\, \xi_{y}(\theta) \exp\left(i{\bm \ell}\cdot{\bm \theta}\right) \nonumber \\
&=&
\int {\rm d}^{2}\theta\, {\cal X}(\xi_{\cal K}(\theta)) \exp\left(i{\bm \ell}\cdot{\bm \theta}\right), \label{eq:power_y}
\eeqa
where the two-point correlation function of $\cal K$ is defined as
\beqa
\xi_{\cal K}(\theta) 
= \int \frac{{\rm d}^2 \ell}{(2\pi)^2}\, \left|\tilde{U}(\ell)\right|^2 P_{\kappa} (\ell). \label{eq:xi_kappa_smoothed}
\eeqa
Hence, for a given PDF and power spectrum of ${\cal K}_{\rm cosmo}$ field,
one can specify the full statistical property of a new Gaussian field $y$
by solving Eqs.~(\ref{eq:set_local_relation}), (\ref{eq:xi_relation})
and (\ref{eq:power_y}).

\subsubsection*{Advantage and limitation}
We should note the advantage and 
the limitation of local-Gaussainised transformation.
Since the one-point PDF and the power spectrum of $\cal K$
can be reproduced by construction, the prediction by local-Gaussainised transformation
can include the appropriate information of ${\cal K}$ and {\it any} 
order of moments $\langle {\cal K}^{\alpha}\rangle$ defined by
\beqa
\langle {\cal K}^{\alpha}\rangle \equiv 
\int_{-\infty}^{\infty}\, {\rm d}{\cal K}_{\rm cosmo}\, 
{\cal K}_{\rm cosmo}^{\alpha}\, {\rm Prob}({\cal K}_{\rm cosmo}). 
\eeqa
Hence, the non-Gaussian information about the one-point PDF
(e.g., skewness $\langle {\cal K}^{3}\rangle$)
is properly taken into account in local-Gaussianized model.
Unfortunately, there is no guarantee that local-Gaussainised transformation
can explain the multi-point clustering beyond power spectrum in ${\cal K}$,
e.g., bispectrum.
The objective of this paper is to examine 
if the correct information of one-point {\it and} two-point PDF of $\cal K$
is sufficient to predict the weak-lensing peak counts.
Local-Gaussianized transformation is suitable for this purpose.

\subsubsection*{Analytic expression of peak counts}
From Eq.~(\ref{eq:local_transform}), one obtains
\beqa
\eta_{{\rm cosmo}, i} 
&\equiv& \nabla_{i}{\cal K}_{\rm cosmo}
= {\cal F}^{\prime} \nabla_{i}y \nonumber \\
&=& {\cal F}^{\prime} \eta_{y, i}, \label{eq:1st_der_kappa} \\
\zeta_{{\rm cosmo}, ij} 
&\equiv& \nabla_{i}\nabla_{j} {\cal K}_{\rm cosmo}
= {\cal F}^{\prime \prime} (\nabla_{i}y)(\nabla_{j}y)
+ {\cal F}^{\prime} \nabla_{i}\nabla_{j}y \nonumber \\
&=& {\cal F}^{\prime \prime} \eta_{y,i}\eta_{y,j}
+ {\cal F}^{\prime} \zeta_{y, ij}, \label{eq:2nd_der_kappa}
\eeqa
where $^{\prime} = {\rm d}/{\rm d}y$. 
Eqs.~(\ref{eq:1st_der_kappa}) and (\ref{eq:2nd_der_kappa})
show that the eigen values of $-{\bm \zeta}_{\rm cosmo}$
are proportional to that of $-{\bm \zeta}_{y}$ at the position of local maxima where ${\bm \eta}_{\rm cosmo} = 0$ (i.e., 
${\bm \eta}_{y}=0$) holds.
Therefore, the expected number density of local maxima of ${\cal K}_{\rm cosmo}$
under the local transformation of Eq.~(\ref{eq:local_transform})
can be derived in a similar way to Gaussian case.
By using Eq.~(\ref{eq:peak_dens_wonoise}), one can find that
\beqa
\frac{{\rm d}n_{\rm peak}}{{\rm d}{\cal K}_{\rm cosmo}}\biggr|_{{\cal K}_{\rm cosmo}=\alpha}
&=&
\int {\rm d}{\bm D}_{\rm cosmo} \, {\rm Prob}({\bm D}_{\rm cosmo})
\delta_{D}({\cal K}_{\rm cosmo} - \alpha) \, \nonumber \\
&&\times
\left|\det{\bm \zeta}_{\rm cosmo}\right|\,
\delta_{D}({\bm \eta}_{\rm cosmo}) \, {\cal H}({\bm \lambda}_{\rm cosmo}) \nonumber \\
&=& \int {\rm d}{\bm D}_{y} \, {\rm Prob}({\bm D}_{y})
\delta_{D}({\cal F}(y) - \alpha) \, \nonumber \\
&&\times
\left|{\cal F}^{\prime}\, \det{\bm \zeta}_{y}\right|\,
\delta_{D}({\bm \eta}_{y}) \, {\cal H}({\cal F}^{\prime} {\bm \lambda}_{y}) \nonumber \\
&=& \frac{1}{2\pi\theta_{y*}^2}\frac{1}{\sqrt{2\pi}\sigma_{y0}}
\frac{1}{{\cal F}^{\prime}}
\exp\left[-\frac{1}{2}\left(\frac{{\cal F}^{-1}(\alpha)}{\sigma_{y0}}\right)^2\right] \nonumber \\
&&\times
G\left(\Gamma_{y}, \Gamma_{y}\frac{{\cal F}^{-1}(\alpha)}{\sigma_{y0}}\right),\label{eq:peak_LG}
\eeqa
where 
${\bm D}_{y} = \{ y, {\bm \eta}_{y}, {\bm \zeta}_{y} \}$,
$\sigma_{yi}$ is the $i$-th moment of $y$ field,
$\theta_{y*}=\sqrt{2}\sigma_{y1}/\sigma_{y2}$,
$\Gamma_{y}=\sigma_{y1}^2/\sigma_{y0}\sigma_{y2}$,
and the function of $G$ is shown in Eq.~(\ref{eq:peak_function_G}).

\section{NUMERICAL SIMULATION}\label{sec:sim}
\subsection{Ray-tracing simulation}
To study the weak-lensing peak count including 
appropriate nonlinear gravitational growth,
we utilise a large set of ray-tracing simulations.
The ray-tracing simulations are constructed from 
$200$ realizations of N-body simulations 
with box sizes of 240 $h^{-1}\, {\rm Mpc}$ 
on a side. 
N-body simulations have been performed with
the number of particles of $256^3$ for 
the concordance $\Lambda$CDM  model.
In simulations, the following cosmological parameters are assumed:
the matter density $\Omega_{\rm m0}=0.238$, 
the baryon density $\Omega_{b} = 0.042$,
the dark energy density $\Omega_{\Lambda}=1-\Omega_{\rm m0}=0.762$,
the equation of state parameters $w=-1$,
the scalar spectral index $n_s = 0.958$,
the amplitude of curvature perturbations
$A_s=2.35\times10^{-9}$ at $k=0.002\, {\rm Mpc}^{-1}$,
Hubble parameter $h=0.732$,
and the variance of the present-day density
fluctuation in a sphere of radius $8 \, h^{-1}\, {\rm Mpc}$
$\sigma_8 = 0.76$.
We work with single source redshift at
$z_{\rm source}=1$.
From the 400 N-body simulations,
we generate 1000 realizations of $5\times5$ sq.degs
lensing convergence fields (i.e., a total of 25,000 squared degrees) 
with $2048^2$ pixels.
The angular size of each pixel is set to be 0.15 arcmin. 
Details of the ray-tracing simulation are found in \citet{2009ApJ...701..945S}.

Throughout this paper, 
we include galaxy shape noise $e$ in our simulation by adding to the measured shear signal random ellipticities which follow the two-dimensional Gaussian distribution as
\beqa
P(e) = \frac{1}{\pi \sigma_{\rm sn}^2} \exp\left(-\frac{e^2}{\sigma_{\rm sn}^2}\right),
\eeqa
where $e = \sqrt{e_{1}^2+e_{2}^2}$ and 
$\sigma_{\rm sn}^2= \sigma_{e}^2/(n_{\rm gal}\theta_{\rm pix})$
with the pixel size of $\theta_{\rm pix}=0.15$ arcmin.
In this paper, we set $\sigma_{e}=0.4$ and study two different cases
of $n_{\rm gal}=10$ and 30 ${\rm arcmin}^{-1}$.
The former source number density corresponds to the typical value
of the current-generation ground-based imaging survey, while
the latter is for the future imaging surveys.

From simulated lensing shear field $\gamma$, 
we construct the smoothed convergence map as shown 
in Section~\ref{subsec:basics} for three different smoothing scales
$\theta_G=1.2, 2.4$ and $4.8$ arcmin in the absence/presence of shape noise.

\subsection{Local-Gaussianized simulation}

\begin{figure}
\centering
\includegraphics[width=0.8\columnwidth, bb=0 0 476 462]
{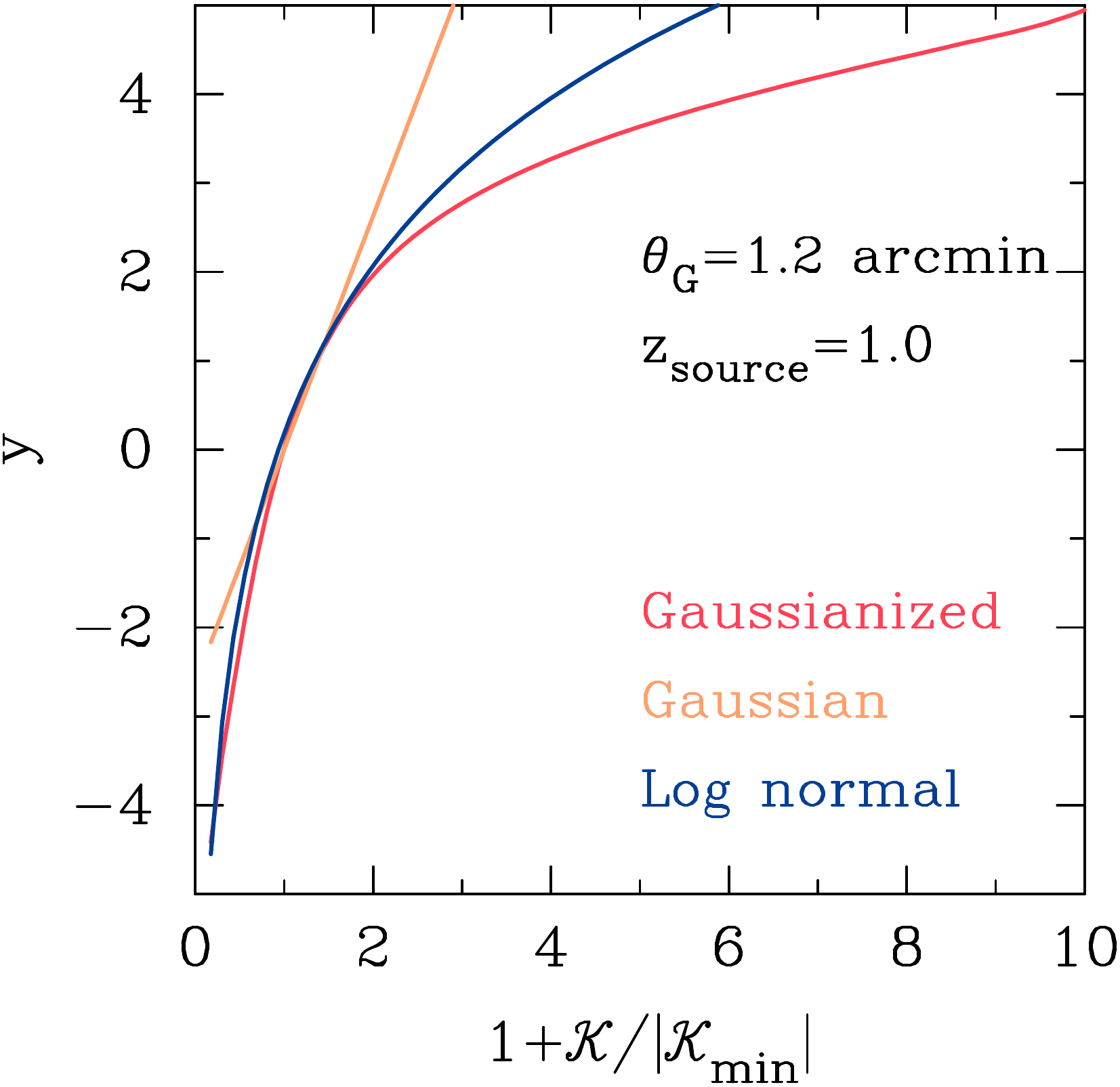}
 \caption{
 An example of local-Gaussianized transformation.
 The red line represents the relation of ${\cal K}={\cal F}(y)$ obtained from the simulated one-point PDF in 1000 ray-tracing simulations.
 The yellow and blue lines show two representative examples:
 the yellow corresponds to the simple linear (Gaussian) relation 
 and the blue is for the log-normal relation.
 In this figure, we set $\theta_G=1.2$ arcmin.
 }
 \label{fig:local_relation}
\end{figure}

In order to specify the local-Gaussianized transformation for 
noiseless $\cal K$ with a given $\theta_G$,
we measure the one-point PDF of $\cal K$ in the absence of shape noise.
We compute the PDF for 90 equally spaced bins of 
$({\cal K}-\langle{\cal K}\rangle)/\sigma_{{\cal K}0}$ 
between -5 to 25 where $\sigma_{{\cal K}0}^2$ is 
the variance of $\cal K$ without shape noise.
Note that we calculate $\sigma_{{\cal K}0}^2$ as the average value 
of the variance over 1000 ray-tracing realizations.
We then define the local transformation to obtain $y$
by using averaged one-point PDF over 1000 ray-tracing 
realizations and Eq.~(\ref{eq:set_local_relation}).
According to Eq.~(\ref{eq:kappa_delta}),
the convergence field should have the minimum value which is given by
\beqa
\kappa_{\rm min} = -\int_{0}^{\chi_{\rm source}} {\rm d}\chi \, W(\chi). \label{eq:kappa_empty}
\eeqa
Nevertheless, the minimum value of smoothed field $\cal K$
would show the dependence on $\theta_{G}$ and does not correspond 
to the value of Eq.~(\ref{eq:kappa_empty}) in practice.
This effect would arise from the finite sampling in the limited survey area
\citep{2002ApJ...571..638T}.
We thus simply extract the minimum ${\cal K}$ from 
1000 ray-tracing realizations
for a given $\theta_G$.
Figure~\ref{fig:local_relation} shows an example of ${\cal K}={\cal F}(y)$ obtained in our analysis for $\theta_G=1.2$ arcmin.
In this figure, we also show two representative examples as
\beqa
1+\frac{\cal K}{|{\cal K}_{\rm min}|}
=
\left\{
\begin{array}{ll}
1+y \, \sigma_{{\cal K}0}/|{\cal K}_{\rm min}|  & ({\rm Gaussian}) \\
\exp \left(-\sigma_{\rm ln}^2/2 + y \, \sigma_{\rm ln}\right) & ({\rm Log \, normal}) \\
\end{array}
\right.
,
\eeqa
where ${\cal K}_{\rm min}$ is the minimum value of simulated $\cal K$
and 
$\sigma_{\rm ln}^2 = \ln(1+\sigma_{{\cal K}0}^2/|{\cal K}_{\rm min}|^2)$ \citep{2002ApJ...571..638T}.
Gaussian expression is a reasonable approximation for $|y|\ll 1$,
while log-normal approximation can capture the feature of 
${\cal K} > {\cal K}_{\rm min}$.
For a long tail in the positive direction of ${\cal K}$,
log-normal approximation is found to be insufficient and this
is consistent with the conclusion of previous works 
\citep[e.g.,][]{2002ApJ...571..638T, 2006ApJ...645....1D, 2009ApJ...691..547W, 2011ApJ...742...15T}.

\begin{figure*}
\centering
\includegraphics[width=1.30\columnwidth, bb=0 0 638 418]
{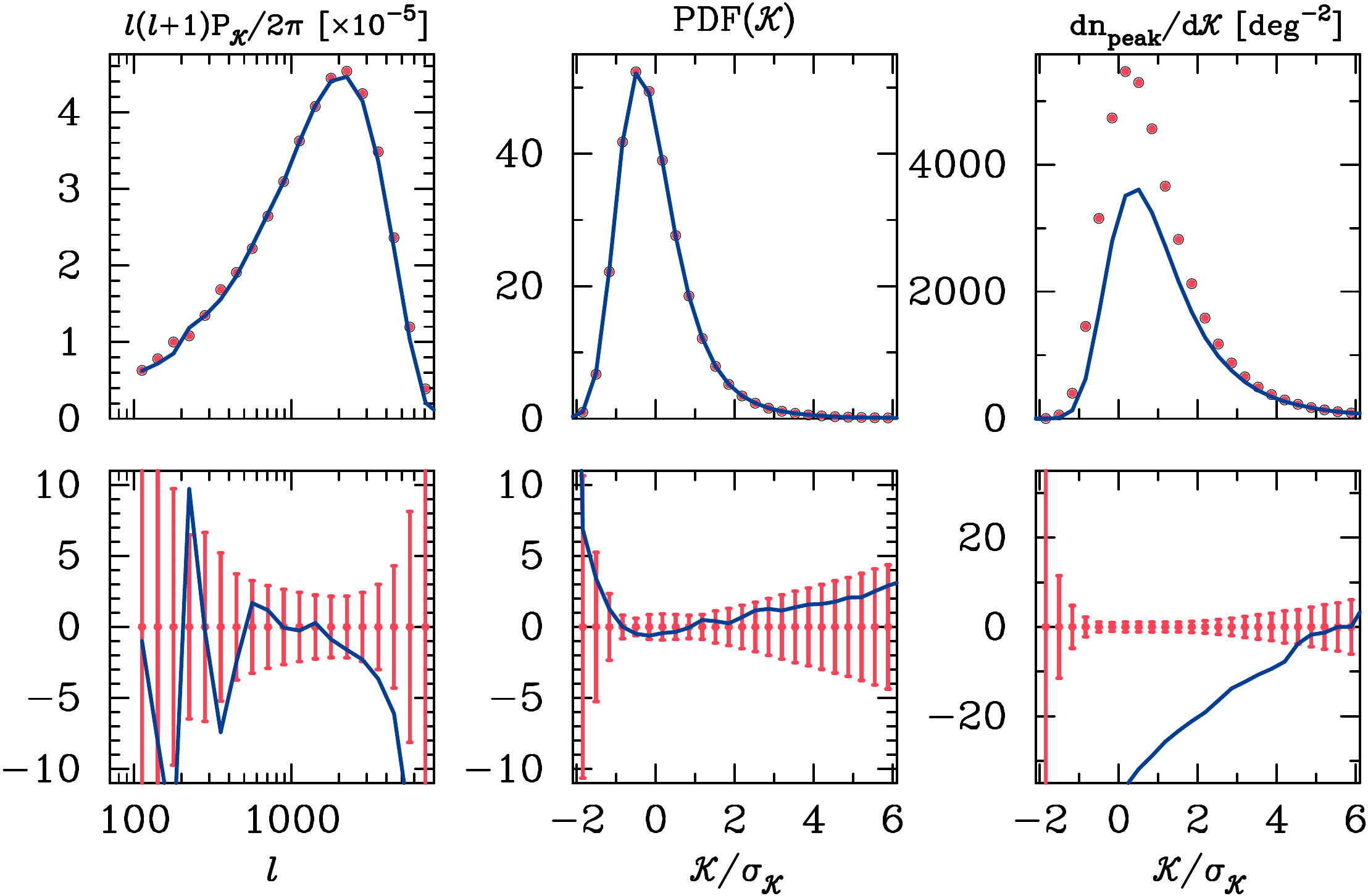}
 \caption{
 Comparison of different statistics in smoothed convergence field
 in the absence of shape noise. In each panel,
 the upper portion shows the comparison of a given statistical quantity 
 measured from 1000 ray-tracing simulations and 1000 local-Gaussianized
 simulations, while the bottom represents the fractional difference in percentage.
 The red points correspond to the case of ray-tracing simulations 
 and the blue line is for the prediction by 
 local-Gaussianized simulations. In the bottom panels, 
 we show the statistical uncertainty for a 1,000 squared-degrees survey,
 which is evaluated by the standard deviation of 1000 simulations 
 with a survey area scaling of $1/{\rm Area}$.
 The smoothing scale is set to be 1.2 arcmin.
 {\it Left}: The power spectrum.
 {\it Middle}: The one-point PDF.
 {\it Right}: The peak counts.
 }
 \label{fig:Pk_pdf_peak}
\end{figure*} 

We also need the power spectrum of ${\cal K}$ 
to find the power spectrum of a new Gaussian field $y$.
We follow the method in \citet{2009ApJ...701..945S} 
to estimate the convergence power spectrum from ray-tracing simulations based on the fast Fourier transform. 
We measure the binned power spectrum of the (un-smoothed)
convergence field 
by averaging the product of Fourier modes $|\tilde{\kappa}(\ell)|^2$. 
We employ 30 bins logarithmically spaced 
in the range of $\ell = 100$ to $5\times10^4$. 
By comparing the theoretical prediction in Eq.~(\ref{eq:kappa_power})
with the fitting formula of $P_{\delta}$ in \citet{2012ApJ...761..152T},
we find that the simulated power spectrum can be approximated as
\beqa
P_{\kappa}(\ell; {\rm simulated})
&\simeq& P_{\kappa}(\ell; {\rm theory}) \left(1.0+\frac{\ell}{7000}\right)^{-0.21}
\nonumber \\
&&\times
\left(1.0+\frac{\ell}{70000}\right)^{-0.12},
\label{eq:Pkappa_sim}
\eeqa
where $P_{\kappa}(\ell; {\rm simulated})$ is the average power spectrum
from 1000 realizations, $P_{\kappa}(\ell; {\rm theory})$ is the theoretical model of Eq.~(\ref{eq:kappa_power}),
and the correction factor represents the resolution effect 
in ray-tracing simulations.
Since our primary focus is to construct the local transformation of Eq.~(\ref{eq:local_transform}) which can reproduce the simulated power spectrum and the one-point PDF of smoothed convergence field,
we work with the power spectrum approximated as in Eq.~(\ref{eq:Pkappa_sim})
to define the local-Gaussianized transformation.
After some trials, we also find that by removing small-$\ell$ modes in  
Eq.~(\ref{eq:Pkappa_sim})  
the variance in ray-tracing simulations is better approximated.
For the input in Eq.~(\ref{eq:power_y}),
we modify the expression in Eq.~(\ref{eq:Pkappa_sim}) as
\beqa
P_{\kappa}(\ell; {\rm input}) = 
\exp\left[-\left(\frac{2\pi}{\theta_{\rm sim}\ell}\right)^2\right]\, P_{\kappa}(\ell; {\rm simulated}), \label{eq:Pkappa_sim_v2}
\eeqa
where 
$\theta_{\rm sim}(=0.15\times2048 \times \pi/180/60\, {\rm rad})$ represents the field of view in ray-tracing simulations.
Note that the difference between Eqs.~(\ref{eq:Pkappa_sim}) and
(\ref{eq:Pkappa_sim_v2}) causes a $\sim10\%$ effect in 
estimation of the variance of ${\cal K}$.
We then define the power spectrum of $y$ by using
Eqs.~(\ref{eq:power_y}) and (\ref{eq:xi_kappa_smoothed})
with the input spectrum of Eq.~(\ref{eq:Pkappa_sim_v2})
for a given $\theta_G$.
Note that we use the fit of convergence power spectrum as in Eq.~(\ref{eq:Pkappa_sim_v2})
not the measured power spectrum from ray-tracing simulations
when computing Eqs.~(\ref{eq:power_y}) and (\ref{eq:xi_kappa_smoothed}).
This is because the binned spectrum from simulations is not suitable to compute
the integral in Eq.~(\ref{eq:xi_kappa_smoothed}).

When we obtain the relation of ${\cal K}={\cal F}(y)$ and the power spectrum $P_{y}(\ell)$, we generate 1000 Gaussian realizations 
with the power spectrum of $P_{y}$
and then transform a Gaussian field $y$ into ${\cal K}$ 
by using ${\cal K}={\cal F}(y)$.
We denote the resulting $\cal K$ field by the local-Gaussianized transformation as ${\cal K}_{\rm LG}$.
In order to simulate the noise convergence field $\cal N$ under 
the local-Gaussianized transformation,
we take the following procedures:
\begin{description}
\item[(i)] create a random Gaussian field
with zero mean and the variance of $\sigma_{\rm sn}^{2}/2$ 
\item[(ii)] perform smoothing with the filter of $U(\theta)$ given by Eq.~(\ref{eq:filter_kappa})
\item[(iii)] add the pixel value found in step (ii) to ${\cal K}_{\rm LG}$
\end{description}
According to these procedures, we can effectively calculate 
the prediction of Eq.~(\ref{eq:peak_dens_wnoise}) 
by counting the local maxima in simulated ${\cal K}_{\rm LG}+{\cal N}$ map (and thus evade the twelve-dimensional integral in Eq.~(\ref{eq:peak_dens_wnoise})).

\section{RESULT}\label{sec:res}

In this section, we compare weak-lensing
peak counts in our simulation with the prediction by local-Gaussianized transformation.
By construction, the theoretical model by local-Gaussianized transformation
should contain proper cosmological information in two-point clustering and any order of moments.
Thus, the comparison shown here is useful 
to clarify how important the information of two-point clustering
and moments is to predict the peak counts.

We first demonstrate how accurate the local-Gaussianized transformation
can predict the one-point PDF and the power spectrum of 
smoothed convergence field $\cal K$ in the absence of shape noise.
In Figure~\ref{fig:Pk_pdf_peak}, the left and middle panels
show the comparison of the power spectrum $P_{\cal K}$
and the one-point PDF, respectively.
The accuracy of local-Gaussianized model is a level of 
$\sim5$\% for $P_{\cal K}$ with the multipole of $\ell\simlt4,000$
and the one-point PDF with 
$-2\simlt{\cal K}/\sigma_{{\cal K}0}\simlt 6$.
On the other hand, we find that 
the local-Gaussianized model can ${\it not}$ predict the peak count
with the similar accuracy to the PDF (or $\sim5\%$).
Interestingly, the largest discrepancy between the simulated peak counts
and the local-Gaussianized model is found at 
${\cal K}/\sigma_{{\cal K}0}\sim1$ 
where the non-Gaussianity in ${\cal K}$ 
is expected to be smaller than the region with 
${\cal K}/\sigma_{{\cal K}0}\simgt 5$.
For high peak heights, the local-Gaussianized model is found to be 
a reasonable expression of simulated peak counts.
It is simply because the peaks with 
${\cal K}/\sigma_{{\cal K}0}\simgt 5$ 
should show a clear one-to-one correspondence to massive dark matter
halos \citep[e.g.,][]{2004MNRAS.350..893H, 2005A&A...442..851M, 2012MNRAS.425.2287H, 2015MNRAS.453.3043S}
and a long tail in the positive direction of $\cal K$
is owing to such halos \citep[e.g.,][]{2000MNRAS.318..321K}.
For peaks with ${\cal K}/\sigma_{{\cal K}0}\simlt-2$,
it is still difficult to obtain the converged result due to small number of peaks.
Note that the minimum value of $\cal K$ is found to 
be $\sim-3\, \sigma_{{\cal K}0}$ in our simulations.

\begin{figure}
\centering
\includegraphics[width=0.80\columnwidth, bb=0 0 460 420]
{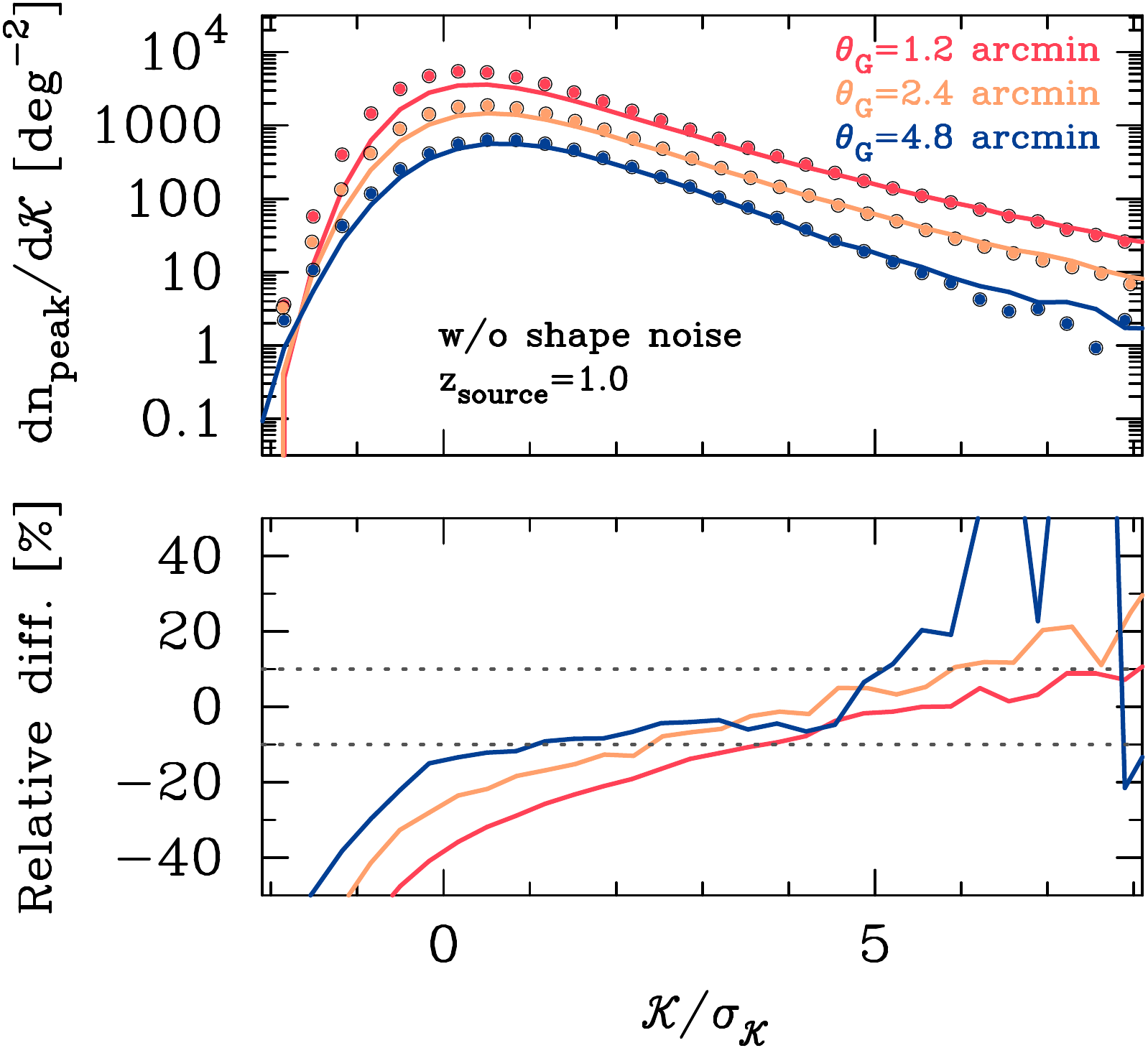}
 \caption{
 Comparions of simulated peak counts with the prediction by 
 local-Gaussianized transformation.
 The coloured points represent the average peak counts from 
 1000 ray-tracing simulations, while the lines are for the prediction
 by 1000 local-Gaussianized simulations.
 The difference of colours corresponds to the difference of smoothing scales.
 The top panel shows the peak count and the bottom shows the fractional differences.
 In this figure, we do not include the shape noise contaminants.
 }
 \label{fig:peak_comp_wonoise}
\end{figure} 

\begin{figure*}
\centering
\includegraphics[width=0.90\columnwidth, bb=0 0 460 420]
{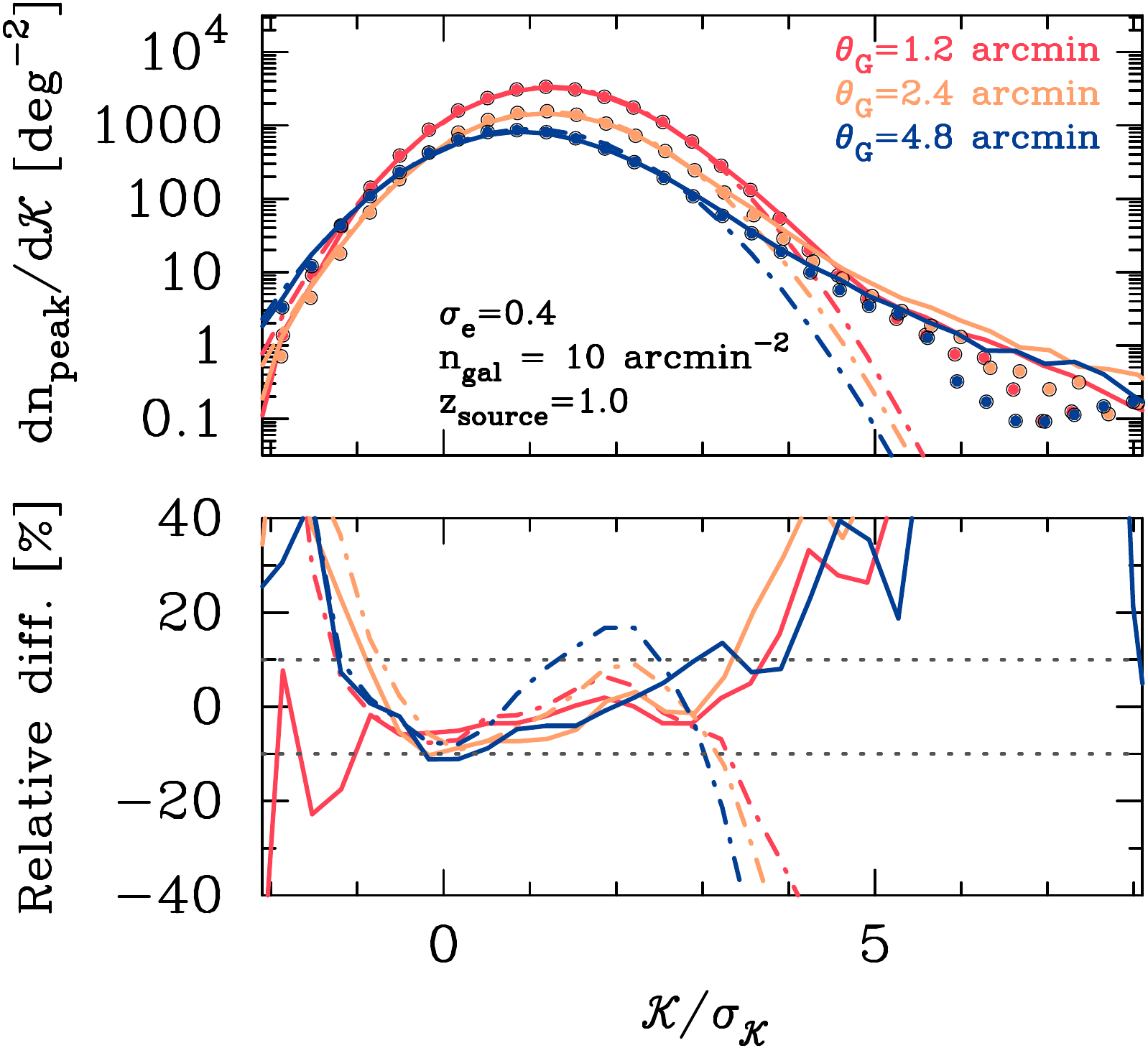}
\includegraphics[width=0.90\columnwidth, bb=0 0 460 420]
{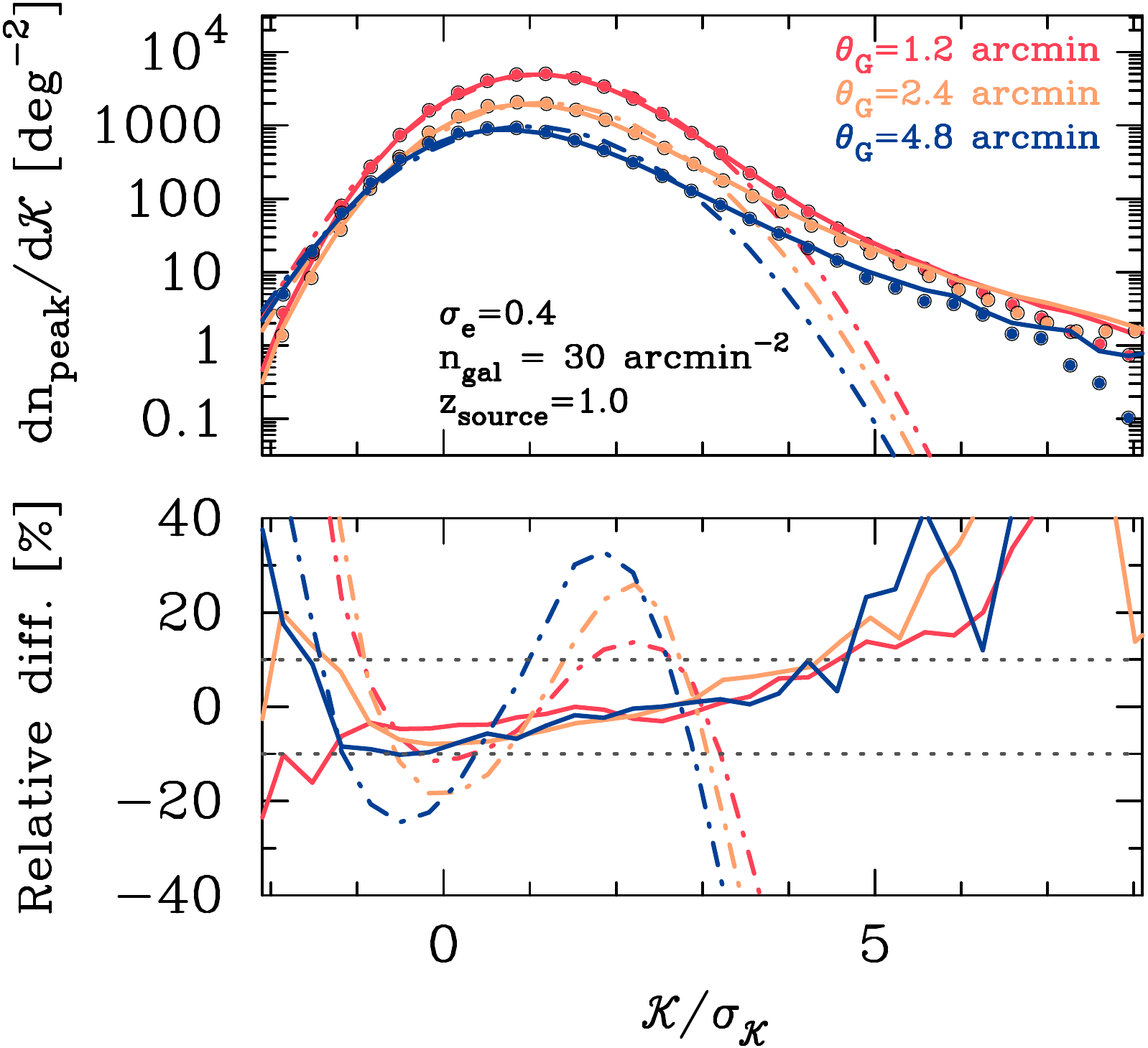}
 \caption{
 Similar to Figure~\ref{fig:peak_comp_wonoise}, but the shape noise
 are taken into account. The left panels summarise the result 
 with the source number density of $n_{\rm gal}=10\, {\rm arcmin}^{-2}$,
 while the right is for $n_{\rm gal}=30\, {\rm arcmin}^{-2}$.
 Note that coloured dashed lines represent the Gaussian prediction 
 for different smoothing scales.
 }
 \label{fig:peak_comp_wnoise}
\end{figure*} 

Figure~\ref{fig:peak_comp_wonoise} shows the comparison of peak counts for different smoothing scales $\theta_G=1.2, 2.4$ and 4.8 arcmin.
Regardless of smoothing scales, we confirm that
the local-Gaussianized model can not explain the simulated peak counts in the range of ${\cal K}/\sigma_{{\cal K}0}\simlt3$. 
The discrepancy between simulated peak counts
and the local-Gaussianized model tends to be smaller when one 
adopt larger smoothing scales.
We carefully calibrate the moment of $\sigma_{{\cal K}i}$
and find that the local-Gaussianized model can predict 
$\sigma_{{\cal K}0}, \sigma_{{\cal K}1}$, and $\sigma_{{\cal K}2}$
with a few percent accuracy.
Thus, the amplitude of peak counts by the local-Gaussianized model 
should be evaluated with the similar accuracy to 
the case of the one-point PDF.
We also find that 
the positive tail in the simulated peak counts can be explained by
the local-Gaussianized model quantitatively.
Although peaks with extremely high height are still rare over 1000 realizations 
and thus require larger simulations to quantify 
the accuracy of local-Gaussianized model in such high $\cal K$ regime,
the local-Gaussianized model can provide a prediction of the peak counts
with ${\cal K}/\sigma_{{\cal K}0}\sim5$ with about 10\% accuracy.

We next consider a more realistic case of weak-lensing peak counts
by including the shape noise in ray-tracing and local-Gaussianized simulations.
Figure~\ref{fig:peak_comp_wnoise} 
shows the comparison of peak counts in the presence of shape noise.
As a reference, the dashed lines represent the Gaussian prediction of
peak counts as in Eq.~(\ref{eq:gauss_peak}).

In the case of the source number density of 
$n_{\rm gal}=10\, {\rm arcmin}^{-2}$ corresponding to
the typical value in the current lensing surveys \citep[e.g.,][]{2013MNRAS.433.2545E}, 
the local-Gaussianized model can predict the simulated peak counts
in the range of $-1\simlt{\cal K}/\sigma_{{\cal K}0}\simlt4$
with a $\sim10$\% accuracy, while the Gaussian prediction
is also found to work within a $10\%$ accuracy 
for $|{\cal K}|/\sigma_{{\cal K}0}\simlt1$.
Note that the current lensing surveys already require 
a prediction of peak counts with an accuracy level of $\sim10\%$ 
\citep[e.g.,][]{2015PhRvD..91f3507L}.
The non-Gaussian tail in simulated peak counts can be also explained 
by the local-Gaussianized model quantitatively even in this realistic case.

When we increase the source number density of $n_{\rm gal}=30\, {\rm arcmin}^{-2}$, 
the Gaussian prediction can not provide an explanation of simulated peak counts even in $|{\cal K}|/\sigma_{{\cal K}0}\simlt1$,
while the local-Gaussianized model can still explain the simulated peak counts with a $\sim10\%$ accuracy
in the range of $-1\simlt{\cal K}/\sigma_{{\cal K}0}\simlt4$.
According to Figure~\ref{fig:peak_comp_wnoise},
the simulated peak counts would surely contain 
the cosmological information of higher-order moments beyond variance
such as skewness even in the presence of shape noise.
Although the local-Gaussianized model can explain the simulated peak counts better than the Gaussian prediction,
there still remains room for improvement on 
theoretical model of weak-lensing peak counts, 
except for including the appropriate information of higher-order moments
than variance such as skewness.

\section{CONCLUSION AND DISCUSSION}\label{sec:con}

We aimed at studying the cosmological information in the number density
of local maxima of weak lensing field or weak-lensing peak counts.
In particular, we have considered the statistical connection to peak counts 
with two-point clustering and moments in weak lensing field.
We first have developed a statistical model of smoothed convergence field 
$\cal K$ assumed to be related with a new Gaussian field $y$.
We refer the relation between $\cal K$ and $y$ as local-Gaussianized transformation.
Our model has been calibrated with 1000 ray-tracing simulations
so that the one-point PDF and the power spectrum of $\cal K$ can be
reproduced with a $\simlt5\%$ accuracy.
Therefore, local-Gaussianized transformation should contain 
the cosmological information of the power spectrum and moments
as similar to the simulated $\cal K$ field.
By comparing the simulated peak counts and the local-Gaussianized prediction, we can assess whether the accurate modeling of 
two-point clustering and any order of moments is sufficient
to predict the peak counts in $\cal K$ field.

We found that the local-Gaussianized transformation is unable
to predict the simulated peak counts in the absence of shape noise
and the differences between two are $\sim30\%$
over a wide range of peak heights for various smoothing scales.
Interestingly, the differences are more prominent for peaks
with the height of $\sim1-2\, \sigma_{{\cal K}0}$
where $\sigma_{{\cal K}0}^2$ represents the variance of 
$\cal K$ field.
These results imply that it is necessary to include other information
beyond two-point clustering and moments in theoretical model of 
peak counts for the first time. 
Also, we studied the effect of shape noise on weak-lensing peak counts
and found that the local-Gaussianized model can provide more reasonable
explanation for simulated peak counts than the Gaussian prediction.
The local-Gaussianized model can achieve a $\sim10\%$
accuracy of weak-lensing peak counts in the presence of shape noise,
whereas the upcoming lensing surveys would require the statistical uncertainty
of a few percent or less.
Hence, we conclude that weak-lensing peak counts
include the cosmological information of higher-order 
moments than second-order one even in the presence of shape noise with a realistic level,
while including the higher-order moments like skewness is not complete
to predict the simulated peak count with the desired accuracy 
for future surveys.
This is consistent with previous theoretical works 
on the inadequacy of multiple-point correlation functions 
to describe nonlinear cosmological fields
\citep[e.g.,][]{2012PhRvL.108g1301C, 2012ApJ...750...28C}.
Our conclusion is also in good agreement with previous lensing studies
showing the moments of the convergence field and of its spatial derivatives
do not contain all the information in the maps 
\citep{2013PhRvD..88l3002P, 2015PhRvD..91j3511P}.
We note that the medium peak contain rich cosmological information even in the presence
of shape noise \citep{2010PhRvD..81d3519K}.

We have shown that the local-Gaussianized transformation can explain
the non-Gaussian tail in the simulated peak counts quantitatively.
As expected, peaks with a significantly positive height are
originated from individual massive dark matter halos along the line of sight.
Such massive halos can also produce a long tail in the positive direction of $\cal K$ 
in the one-point PDF.
Therefore, in high $\cal K$ regime, 
the local mapping ${\cal K}({\bm \theta})$ 
to a Gaussian field $y({\bm \theta})$ 
is a reasonable approximation.
On the other hand, the local mapping does not accurately reproduce 
medium peaks with the height of $\simlt1-2\, \sigma_{{\cal K}0}$ 
because previous works have shown such peaks might be induced
by the superposition of massive dark matter halos along the line of sight
\citep[e.g.,][]{2011PhRvD..84d3529Y, 2016PhRvD..94d3533L}.
Obviously, the smoothing procedure would also make the local mapping
${\cal K}({\bm \theta}) \mapsto y({\bm \theta})$ invalid.
Thus, a non-local relation between $\cal K$ and a Gaussian field $y$
would be needed.
Nonlocality should, by definition, include the information of multiple-point clustering 
in cosmological $\cal K$ field.
In the appendix, we demonstrate the importance of including 
the information on correlation between Fourier modes
in modeling of weak-lensing peak counts.

The numerical study in this paper 
is the first step to clarify the cosmological information of weak-lensing
peak counts and its relation to other statistics. 
Although accurate theoretical model of peak counts needs to be further developed,
our analysis proved that peak counts should contain the cosmological information of two-point clustering, any order of moments, and ${\it others}$ in weak lensing field.

\section*{acknowledgments}
The author thanks an anonymous referee for careful reading 
and suggestion to improve the article.
The author would like to thank Naoki Yoshida for helpful discussions and 
comments on the manuscript. 
The author appreciates the helpful comments of Jia Liu, Takashi Hamana, 
and Zoltan Haiman.
The author is supported by Research Fellowships of the Japan Society 
for the Promotion of Science (JSPS) for Young Scientists.
Numerical computations presented in this paper were in part carried out
on the general-purpose PC farm at Center for Computational Astrophysics,
CfCA, of National Astronomical Observatory of Japan.

\bibliographystyle{mnras}
\bibliography{mn-jour,bibtex}

\appendix
\section{Missing information in local-Gaussianized transformation for modeling of weak-lensing peak counts}

In this appendix, we examine if 
weak-lensing peak counts can be reproduced by
including the correct information of Fourier-mode distribution.
We consider the inverse transformation of Eq.~(\ref{eq:local_transform})
and apply it to 1000 ray-tracing simulations in order to obtain 
its counter-part of Gaussian field $y$:
\beqa
y_{{\rm RT}}({\bm \theta}) 
= {\cal F}^{-1}({\cal K}_{{\rm RT}}({\bm \theta})),
\eeqa
where ${\cal K}_{{\rm RT}}$ represents the smoothed convergence field
obtained from ray-tracing simulation.
Under the local transformation of Eq.~(\ref{eq:local_transform}),
the statistical information of $\cal K$ field in ray-tracing simulations should
be preserved in that of $y_{\rm RT}$.
The full statistical information of $y_{\rm RT}$ can be characterised by 
the distribution in Fourier mode $y_{{\rm RT}, {\bm \ell}}$.
In general, the distribution of Fourier mode $A_{\bm \ell}$ 
is expressed as the function of norm $|A_{\bm \ell}|$
and phase $\theta_{A, {\bm \ell}}$.
For a Gaussian field, the distribution of $A_{\bm \ell}$ is given by
\beqa
{\rm Prob}(|A_{\bm \ell}|, \theta_{A, {\bm \ell}})
&\equiv& {\cal P}_{G}(|A_{\bm \ell}|) {\cal P}_{G}(\theta_{A, \bm \ell}), \label{eq:gauss_fourier} \\
{\cal P}_{G}(|A_{\bm \ell}|) 
&=& 
\frac{2}{\sqrt{P_{A}(\ell)}} \left(\frac{|A_{\bm \ell}|}{\sqrt{P_{A}(\ell)}}\right) \nonumber \\
&&\times
\exp\left[-\left(\frac{|A_{\bm \ell}|}{\sqrt{P_{A}(\ell)}}\right)^2\right], \label{eq:gauss_fourier_norm} \\
{\cal P}_{G}(\theta_{A, {\bm \ell}}) 
&=& 
\frac{1}{2\pi}, \label{eq:gauss_fourier_phase}
\eeqa
where $P_{A}(\ell)$ is the power spectrum of $A$.
For a weakly non-Gaussian field, the perturbative approaches have been
developed \citep{2003ApJ...591L..79M, 2007ApJS..170....1M} 
and the lowest-order non-Gaussian effect in ${\rm Prob}(|A_{\bm \ell}|, \theta_{A, {\bm \ell}})$ is related to 
the three-point correlation in Fourier space, or Bispectrum.
Also, the correlation between norm $|A_{\bm \ell}|$ 
and phase $\theta_{A, {\bm \ell}}$ are naturally
induced in the non-Gaussian field.

In our local-Gaussianized model, we distribute the Fourier mode of $y$ 
with Eq.~(\ref{eq:gauss_fourier}).
Here we extend the model in two different ways
by directly using the distribution of $y_{{\rm RT}, {\bm \ell}}$ 
constructed from ray-tracing simulations.
In the first model, we keep the distribution of $|y_{\bm \ell}|$ as in Gaussian with 
Eq.~(\ref{eq:gauss_fourier_norm})
but include the phase correlation with 
the distribution of $\theta_{y{\rm RT}, {\bm \ell}}$.
We denote this as phase-shared model. 
In the phase-shared model, we realise the similar morphology in 
${\cal K}$ to a given ray-tracing map ${\cal K}_{\rm RT}$
since we use the same distribution of the Fourier phase 
as ray-tracing simulations.
In the second model, we distribute the Fourier phase of $y$ randomly as in 
Eq.~(\ref{eq:gauss_fourier_phase}), while we use the simulated distribution
of $|y_{{\rm RT}, {\bm \ell}}|$ instead of Eq.~(\ref{eq:gauss_fourier_norm}).
We denote this model as norm-shared model.
In phase- and norm-shared models, we include 
the multiple-point correlations of Fourier-mode phase and norm, respectively.
Although we can include {\it any} order of multiple-point correlations of 
phases (norm) in phase-shared (norm-shared) models, we ignore the correlation {\it between} the Fourier-mode phase and norm.
For both phase- and norm-shared models, we generate 1000 realizations of $y$ field and then convert $y$ into $\cal K$ with 
Eq.~(\ref{eq:local_transform}).

\begin{figure}
\centering
\includegraphics[width=0.85\columnwidth, bb=0 0 481 504]
{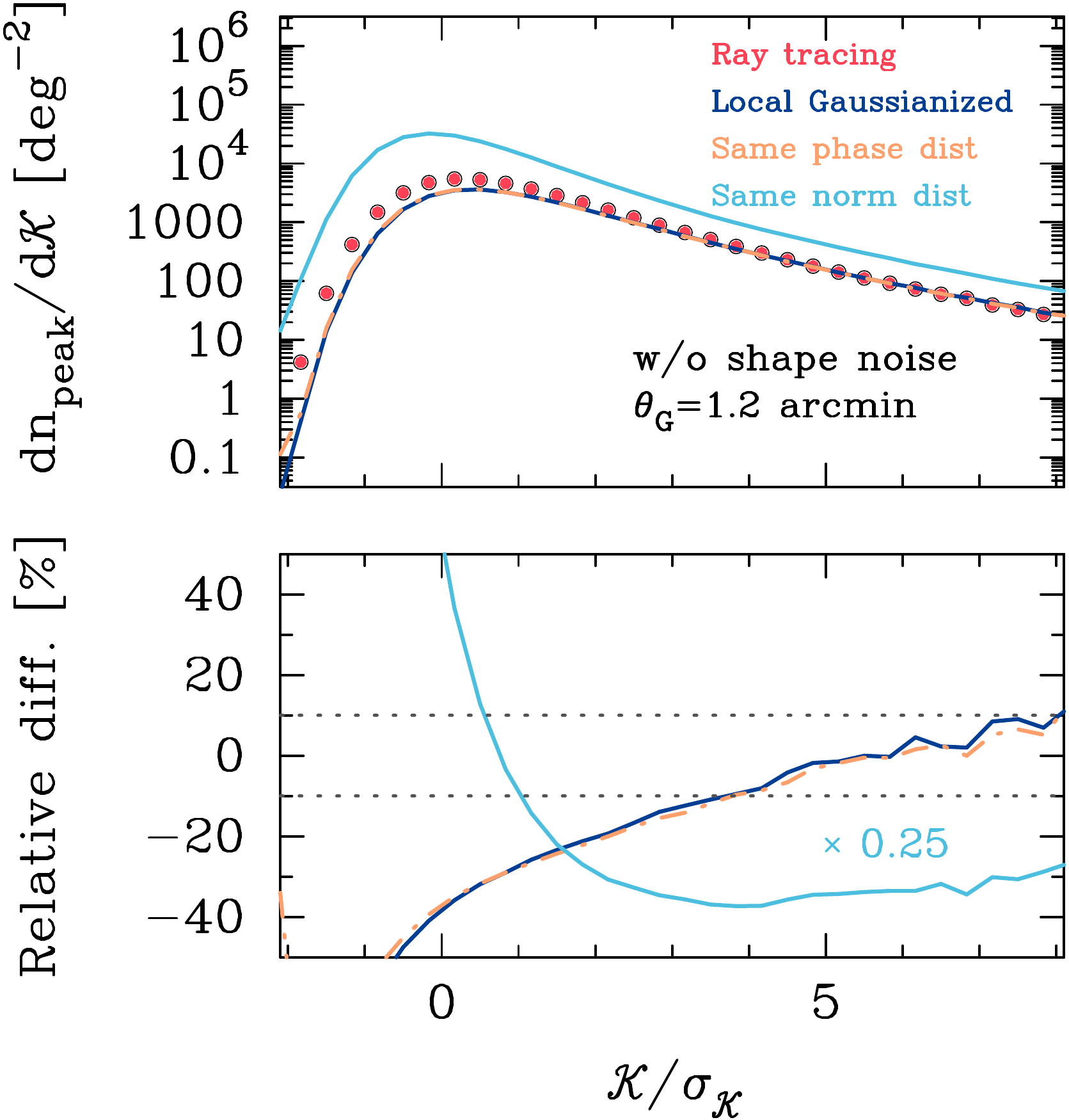}
 \caption{
 Connection of weak-lensing peak counts with the Fourier-mode statistics.
 Figure representation is similar to Figure~\ref{fig:peak_comp_wonoise}.
 The red points represent the average peak counts over 1000 ray-tracing simulations.
 The different coloured lines correspond to the theoretical models based on 
 a local transformation of lensing field ${\cal K}$ to a new field $y$.
 The local relation between $\cal K$ and $y$ is defined by Eq.~(\ref{eq:set_local_relation}), while the statistical property of $y$ field
 in Fourier space is different among them.
 The blue line is our baseline model assumed that $y$ follows Gaussian.
 The yellow line is similar to the blue one but takes into account the phase correlation, while the cyan line shows a model with the correct distribution of the amplitude of Fourier mode $|y_{{\bm \ell}}|$ and the random phase. The details are found in the text. 
 In the bottom panel, we scale the cyan line by a factor of 0.25 
 for illustrative purpose.
 }
 \label{fig:peak_comp_wonoise_share_info}
\end{figure}

Figure~\ref{fig:peak_comp_wonoise_share_info}
shows the comparison of peak counts among the different models.
The red points represent the simulated peak counts measured in 1000 
ray-tracing simulations.
The blue, yellow and cyan lines correspond 
to the local-Gaussianized model, the phase-shared model
and the norm-shared model, respectively.
In this figure, we ignore the shape noise contaminants 
and set the smoothing scale to be 1.2 arcmin.
We find that the phase correlation does not improve the model prediction 
by local-Gaussainised transformation.
The effect of the phase correlation on peak counts 
is found to be $\simlt1\%$.
This indicates that the correct morphological information in $\cal K$ 
would be of little importance to predict the weak-lensing peak counts.
We also find that the information of multiple-point correlation of the Fourier-mode norm significantly increases the number of peaks over
a wide range of peak heights by a factor of $3-5$, 
while the variance of $\cal K$ in the norm-shared model 
is found to be consistent with full ray-tracing simulations with a level 
of $\sim30\%$.
Also, the difference between the norm-shared model
and the ray-tracing result depends on peak height,
showing that a simple correction by 
scaling of the overall amplitude of peak counts can not work.

Hence, the difference of weak-lensing peak counts between ray-tracing simulations and 
our local-Gaussainised model are caused by
the distribution of $y_{\rm RT}$ in Fourier space.
In particular, the phase correlation in $\cal K$ would be minor,
while the correlation between the Fourier-mode norm 
$|y_{\bm \ell}|$ and phase $\theta_{y, {\bm \ell}}$
should play a central role to determine the cosmological information 
of weak-lensing peak counts.

\end{document}